\documentclass[12pt, english]{article}
\usepackage[latin1]{inputenc}
\usepackage{graphics}
\usepackage{epsfig}
\usepackage{amsmath}
\usepackage{amssymb}
\usepackage{latexsym}
\usepackage{cmmib57}
\usepackage{amsfonts}

\newcommand{\be}{\begin{equation}}
\newcommand{\ee}{\end{equation}}
\newcommand{\vse}{\vspace{0.01cm}}
\newcommand{\nsp}{\!\!\!}

\begin{document}
\hyphenation{coun-ter-terms}
\hyphenation{re-gu-la-ri-za-tion}
\hyphenation{e-xa-ctly}
\hspace{4cm}

\begin{center}
{\bf THE CASIMIR EFFECT FOR PARALLEL PLATES REVISITED}
\end{center}

\vspace{1.5cm}

\begin{center}
  {\sc N. A. Kawakami\footnote{supported by FAPEMIG, kawakami@fisica.ufmg.br} and M. C. Nemes\footnote{carolina@fisica.ufmg.br}} \\
  Departamento de Física, \\[-0.1cm]
  Instituto de Ciências Exatas, Universidade Federal de Minas Gerais, \\[-0.1cm]
  Caixa Postal 970, CEP 30161-970, Belo Horizonte, Minas Gerais, Brazil \\[0.5cm]

  {\sc Walter F. Wreszinski\footnote{supported in part by CNPQ, wreszins@fma.if.usp.br}} \\
  Departamento de Física Matemática,\\[-0.1cm]
  Instituto de Física,
  Universidade de São Paulo, \\[-0.1cm]
  Caixa Postal 66318 - 05315-970 São Paulo - Brazil

\end{center}

\vspace{3cm}

{\small{\noindent {\sc Abstract} --- The Casimir effect for a
massless scalar field with Dirichlet and periodic boundary conditions (b.c.) on  infinite parallel plates is revisited in the local quantum field theory (lqft) framework introduced by B.Kay. The model displays a number of more realistic features than the ones he treated. In addition to local observables, as the energy density, we propose to consider intensive variables, such as the energy per unit area $\varepsilon $, as fundamental observables. Adopting this view, lqft rejects Dirichlet (the same result may be proved for Neumann or mixed) b.c., and accepts periodic b.c.: in the former case $\varepsilon $ diverges, in the latter it is finite, as is shown by an expression for the local energy density obtained from lqft through the use of the Poisson summation formula. Another way to see this uses methods from the Euler summation formula: in the proof of regularization independence of the energy per unit area, a regularization-dependent surface term arises upon use of Dirichlet b.c., but not periodic b.c.. For the conformally invariant scalar quantum field, this surface term is absent, due to the condition of zero trace of the energy momentum tensor, as remarked by B.De Witt. The latter property does not hold in the application to the dark energy problem in Cosmology, in which we argue that periodic b.c. might play a distinguished role.

\noindent {\bf 0. \ Introduction and Summary}

\vspace{0.3cm}

The Casimir pressure of the electromagnetic field enclosed by
(infinitely thin) parallel plates, measured by Spaarnay, is one of
the most famous objects in quantum field theory (and in quantum optics)
\cite{lamoreaux}, \cite{mostepanenko}. A good up-to-date account, including recent experiments, is to be found in the volume Séminaire Poincaré, vol.1, with the title ``Énergie du Vide''.

In spite of the well-known exact solution for the pressure
(\cite{lamoreaux}, \cite{mostepanenko}), the energy per unit area
appeared to remain divergent, due to the (nonintegrable) divergence
of the energy density at the boundaries -- a phenomenon analyzed
quite generally in the pioneering paper of Deutsch and Candelas \cite{deutsch} --
until recent work by H. Kühn clarified the situation, showing that the divergences due to
the electric-field and magnetic-field components exactly cancel \cite{kuhn}. This
argument does not, however, hold for other fields, which may play a role in
Cosmology, particulary in the problem of dark energy (\cite{godlowski}, \cite{wreszinski}, \cite{turner}).
There, the energy density -- the (time)-00 component of the energy-momentum tensor $  T_{00} ( x )  $ -- is as important
an observable as the pressure. If it diverges, or is ill-defined, as in the case of general fields
enclosed by parallel plates, the situation remains highly unsatisfactory from a conceptual point of view. In fact,
this is the most elementary example of cutoff-dependence of (in principle) observable quantities, which has been
emphasized by Hagen in more general situations \cite{hagen}.

In this paper we revisit the Casimir effect for a massless scalar field with Dirichlet and periodic boundary conditions (b.c.)
on (infinitely thin) parallel plates. Other b.c. (Neumann or mixed) may also be handled by the same methods,
and yield results qualitatively similar to the Dirichlet case.

In section 1 we introduce the general framework and ideas, which go back to B.S. Kay \cite{kay} and L. Manzoni, G. Scharf
and one of us (W. Wreszinski) (\cite{scharf}, \cite{manzoni}). (See
also the paper by Hollands and Wald for a very stimulating
introduction to the local quantum field theoretic (lqft) aspects of the problem \cite{hollands}.)  There we show that lqft yields a definite formula (1.27) for the energy-density operator.

In section 2, we prove that  closed form expressions (2.12a,b) for the true vacuum's ( i.e., in the presence of the plates) energy density follow from (1.27). This is the answer provided by lqft to the problem. For Dirichlet b.c. this energy density (2.12a) displays, however, a nonintegrable divergence at the plates, thereby yielding (nonphysical) infinite values for two fundamental observables, the energy per unit area and the pressure. The renormalized one-point function is, moreover, also given by a nonintegrable function and , thus, does not define a (Schwartz) distribution. In contrast, for periodic b.c., the energy per unit area is finite and homogeneous by (2.12b). The reason for the  infinity in the Dirichlet case is that only for two space-time
dimensions is  the renormalization assumption (1.5) powerful enough to
yield a divergence-free theory \cite{kay}. Otherwise, additional divergences arise, due to the sharpness of the surface, as remarked in the pioneering paper of
Deutsch and Candelas \cite{deutsch}. The problem occurs whenever the
attempt is made to impose b.c. on quantum fields (called ``unnatural acts'' by R. L. Jaffe in \cite{jaffe}), i.e., to restrict quantum fields to sharp surfaces.

In a more specific lqft context, a similar problem arises in the
restriction to a causal surface (the horizon) in connection with the
problem of localization entropy \cite{schroer}. The effect observed in
\cite{schroer} is, as remarked there, the thermal counterpart of
Heisenberg's old observation that, in approaching conserved global
charges as spatial limits of integrated charge densities, one must
control the vacuum fluctuation in the neighbourhood of the
boundary. The divergence of the inverse ``split distance'' in
\cite{schroer} has, as also remarked there, nothing to do with an
ultraviolet divergence. Both features have strong analogy to the
results in section 2. There, the ultraviolet cutoff drops out from the
calculation: it is merely due to the introduction of the
(cutoff-dependent) regularized fields (1.22) as an intermediate step,
which seems to be unavoidable due to the subtle cancellations
occurring in (1.27), and was also the route followed by B.Kay in
\cite{kay}, who added that ``experience in quantum field theory leads us
to expect that, after renormalization, results will be
regularization-independent for a wide class of regularization
procedures.''

In this paper we are able to develop this conjecture, thereby improving
on \cite{kay} in this aspect. This turns out to be a nontrivial task,
but it provides an additional, important information: we prove
in section 3 that the above-mentioned regularization independence (RI)
within a wide class implies - in a purely mathematical context, of
a rigorous asymptotic analysis along the lines of the Euler-Maclaurin
summation formula \cite{hardy} - the existence of a \textbf{
regularization-dependent surface term} (RDST) (the second term in (3.17a))
which diverges as the cutoff is removed. This very nonlocal term is,
as remarked, due to sharpness of the (infinitely thin) surface in the case of Dirichlet b.c..

The RDST has a long history, both in the physical and in the mathematical literature. In the physical literature, it is associated with the names of Symanzik \cite{symanzik}, Candelas [33],Deutsch and Candelas \cite{deutsch},and Barton [34]. in section 4 we call them, for brevity, Symanzik counterterms. in the mathematical literature, the surface term is well-known as the area term in Weyl's formula, see, e.g., [35]. However, this is not intrinsic to lqft, but depends on the regularizer (which is the exponential one (2.1) in [35]). The introduction of a regularizer as an intermediate step seems to be unavoidable because of the subtle cancellations occurring in (1.27). The regularized version of (1.27) becomes (1.28), but, as seen, the regularizer drops off in section 2.

On the other hand, Elizalde \cite{elizalde} has shown that Hadamard regularization of the integrals occurring in the models treated in \cite{jaffe} yields terms which, in the limit of a sharp surface, remain finite. In section 4 we show that the same happens in the present model. This is accomplished by the use of the Poisson summation formula, whose importance is even better appreciated in the massive case, where logarithmic divergences occur [36]. The interest of the Hadamard regularization is two-fold: it is, as Elizalde remarks, able to separate and identify the singularities as physically meaningful cutoffs associated to a finite width (see (4.6)), on the other hand the corresponding one-point function (which is a constant for periodic b.c.) is a pseudo-function with the same type of singularity with respect to the singular surfaces as the (old) Hadamard form (5.3) of the \textbf{two point} function. The latter has a fundamental meaning in locally covariant quantum field theory \cite{brunetti}, and it seems of interest to inquire upon the eventual significance of this structure in the present context, see section 5, where we also present our conclusions.

We end this introduction with some important remarks concerning the physical motivation of the present paper. The RDST are identical to those surface terms mentioned by B.S. de Witt in his elegant analysis [37](p.307)(see also L.H.Ford [39]). As he remarks, similarly to the electromagnetic case, for a \textbf{ conformally invariant} massless scalar field, with traceless stress tensor, these surface terms vanish! The question may thus well be asked whether this is not a better and more fundamental proposal to get rid of the bothersome (divergent) surface terms, why is the scalar field of any interest at all. Perhaps the infinity just reflects the fact that the massless scalar field is an unphysical object, at least in the context of the Casimir effect. 

We now argue that this is not so, at least as regards the possible relevance of the Casimir effect to the dark energy problem in Cosmology, for which there is both theoretical \cite{turner} and experimental \cite{godlowski} evidence. This is our main physical motivation; see \cite{hollands} for an introduction to the subject. Of course, the geometry of the present model is not relevant to these (potential) applications: an adequate choice would be a closed surface involving a point, e.g., a cube. The parallel plates offer, however, a much simpler case in which the Casimir problem for a scalar field may be studied rigorously.

The first point is that the stress tensor is \textbf{not} supposed to be traceless, being of the form
$$
T_{\mu\nu} = g_{\mu\nu} \rho
\eqno{\rm{(0.1)}}
$$
where $g_{\mu\nu}\equiv (1,-1,-1,-1)$ for flat (Minkowski) space-time - a good approximation for the present cosmic time \cite{[38]}. Above, $\rho$ denotes the energy density. 
We now come to the second (related) point. From (0.1),
$$
p = -\rho
\eqno{\rm{(0.2)}}
$$
where $p$ is the pressure. Taking for $\rho$ the dark energy density, and $p$ the  pressure exerted by dark energy, (0.2) seems to be very well satisfied by the data for dark energy \cite{turner}. For a conformally invariant massless scalar field,
$$
\rho < 0
\eqno{\rm{(0.3)}}
$$
\cite{[37]}. Assumption (0.3) would contradict, together with (0.2) the fact that the observed pressure is \textbf{negative}, which is, perhaps, the most surprising property of dark energy, and the one which led Turner \cite{turner} to declare that ``vacuum energy is almost the perfect candidate for dark energy''. (The argument of [46], leading to the result that the energy density is nonpositive in quantum field theories, is not (generally) valid in the Casimir effect, because in [46] covariance of the fields with respect to the Euclidean group of translations is used in the proof.) It is, in fact, precisely the negative pressure that is difficult to obtain in alternative proposals for dark energy (see also [40]).(0.3) holds for the model treated in \cite{kay}. By (3.17a), it also holds for the finite part of the energy density of the massless  scalar field, but the latter is positive for the massive scalar field for suitable values of the mass in the case of parallel plates [36], as well as for the inner problem of the cube \cite{wreszinski}. As an example of a massive scalar field , one may consider the hypothetical axion, which may have very small nonzero mass [41]. The latter is a candidate for cold dark matter \cite{turner}, and one may investigate its corresponding vacuum field. Of course, a massive scalar field has no conformally invariant counterpart, and the surface terms do occur in this case (except for periodic b.c., see section 5). We postpone further comments on this and related issues to the conclusion in section 5.
We refer to \cite{lamoreaux}, \cite{mostepanenko} and \cite{elizalde} for (part of) the immense literature on the Casimir effect. For the applications to Gravitation and Cosmology, there is an early beautiful review by L. Parker [45], as well as recent papers by Milton, Elizalde and coworkers [42], [43], [44], where many other references can be found.

\vspace{0.7cm}

\noindent {\bf 1. \ General Framework}

\vspace{0.3cm}

We consider a massless scalar field $\Phi (x)$ on Minkowski space time $ x \equiv  (  x_{0} , \vec{x} )  $. The
corresponding Hamiltonian density $h(x)$ is given by $(\hbar=c=1)$:

$$
h(x)= \frac{1}{2} \Bigg[\bigg(\frac{\partial \Phi (x)}{\partial x _0}\bigg)^{2} + \big(\nabla \Phi (x)\big)^{2}\Bigg]
\eqno{\rm{(1.1)}}
$$
\vse

We also wish to consider the free (massless scalar) field restricted to the region $K _{d}$ between two (infinitely
thin) parallel plates at $z=0 \textrm{ and } z=d$:

$$
K_{d}=\Big \{ \vec{x} \equiv (\vec{x} _{\parallel},z), \textrm{ with } \vec{x} _{\parallel} \equiv (x,y) \in \mathbb{R}^{2}
\textrm{ and } 0 \le z \le d \Big \}
\eqno{\rm{(1.2)}}
$$
\vse

\noindent with Dirichlet or periodic boundary conditions (b.c.) on the boundary
$\partial K_{d}$ of $K_{d}$:

$$
\partial K _{d} \equiv \Big \{ (\vec{x} _{\parallel}, 0) \cup (\vec{x} _{\parallel}, d); \vec{x} _{\parallel}
\in \mathbb{R}^{2} \Big \}
\eqno{\rm{(1.3)}}
$$
\vse

The density operator corresponding to (1.1) is given by

$$
H(x) = :h(x):
\eqno{\rm{(1.4)}}
$$
\vse

\noindent where the dots indicate normal (or Wick) ordering. Measuring $H(x)$ is a local operation which involves only
a small neighborhood $N(x)$ of the space-time point $x$. Since, however, the \underline{state} $S$ of the system on
$K_{d}$ is different from the vacuum state $\omega$ of (infinite) space-time even restricted to $N(x)$ (see \cite{kay},
appendix, for a discussion), the question arises: with respect to which state is the normal ordering (1.4)? In \cite{kay},
the following \underline{renormalization condition} was imposed:

$$
\omega (H(x))=0
\eqno{\rm{(1.5)}}
$$
\vse

\noindent for all $x$ in Minkowski space-time. This condition means that double dots refer to the infinite-space Minkowski
vacuum state $\omega$, and was motivated in \cite{scharf}, \cite{manzoni} by the fact that real boundaries consist of
electrons and ions, and the field which interacts with them is quantized in infinite space, but one may also view (1.5)
as an independent renormalization condition, as done by B. S. Kay in \cite{kay}. The assumptions of local quantum theory
\cite{haag} yield now a rigorous formula for $S(H(x))$ (see, again, the appendix of \cite{kay}):

$$
S(H(x))\; =\;\lim_{x_1,x_2\to x}\;\;\frac{1}{2} \,\Big(\partial_{x_{01}}
     \partial_{x_{02}} + \partial_{\vec{x}_1}
      \partial_{\vec{x}_2}) \omega_{1,+}(x_1,x_2)
\eqno{\rm{(1.6a)}}
$$
where
$$
\omega_{1,+}(x_1,x_2)=\;
 S(\Phi(x_1) \Phi(x_2))
        -\omega (\Phi(x_1) \Phi(x_2)) 
\eqno{\rm{(1.6b)}}
$$
\vse

The scalar field of zero mass, quantized in infinite space in $p$ dimensions, may be formally written

$$
\Phi^{(p)} (x)=
\frac{1}{(2\pi)^{p/2}}\,\int \!\!\frac{d^p k}{\sqrt{2\omega_{\vec{k}}}}
\,\left[a(\vec{k}\,) e^{-ik\cdot x} + a^+ (\vec{k}\,) e^ {ik\cdot
    x}\right] = \Phi_-^{(p)} (x) + \Phi_+^{(p)} (x) \eqno{\rm{(1.7)}}
$$
\vse

where $ x\equiv (x_0, \vec{x}\,), k\cdot x = k_0 x_0 - \vec{k}
\cdot \vec{x}$, $$
k_0 = \omega_{\vec{k}} = |\vec{k}|
\eqno{\rm{(1.8)}} $$
\vse
and $\,\Phi_- , \Phi_+\,$ refer to the negative
and positive-frequency parts in (1.7), i.e., those associated to a
(resp. $a^+$), and satisfy $$
\left[\Phi_-^{(p)} (x), \Phi_+^{(p)}
  (y)\right] = \frac{1}{i} \, {D_{0,p}^{(+)}} (x-y)\;.
\eqno{\rm{(1.9)}} $$
\vse

\noindent with
$$
{D_{0,p}^{(+)}} (x) \equiv
\frac{1}{(2\pi)^p}\, \int \!\!\frac{d^p k}{2\omega_{\vec{k}}}\,\;
e^{-ik\cdot x} \eqno{\rm{(1.10)}} $$
\vse

\noindent $a_{p}$, $a^{+}_{p}$ are annihilation and creation operators defined on symmetric Fock space over
the (one-particle) Hilbert space $\mathcal{H}_{p}=L^2(\mathbb{R}^p)$, $\mathcal{F}_s(\mathcal{H}_{p})$, with
$a(f)$ antilinear, $a^{+}(g)$ linear, such that

$$
\Big [a_{p}(f), a^{+}_{p}(g) \Big ] = (f,g)_{\mathcal{H}_{p}} \mathbf{1}
\eqno{\rm{(1.11a)}}
$$
\vse

\noindent on a dense domain $\mathcal{F}_{0}$ of finite-particle vectors (see, e.g., \cite{reed}, p.208 ff.),
where $(f,g)_{\mathcal{H}_{p}}$ denotes the scalar product on $\mathcal{H}_{p}$. The vacuum $\Omega _{p}$ is
such that

$$
a_{p}(f) \Omega_{p}=0 \;\;\;\; \forall f \; \in \;\ \mathcal{H}_{p}
\eqno{\rm{(1.11b)}}
$$
\vse

The scalar field of zero mass on the region $K_{d}$ is formally given by

$$
\Phi_{K_{d}}(x)=\frac{1}{(2\pi)^2}\, \int \!\!d\vec{k}_{\parallel} \sum _{n=1}^{\infty}
\frac{1}{\sqrt{2\omega_{\vec{k}_{\parallel,n}}}}
\Big (a(\vec{k},n)U_{\vec{k}_{\parallel,n}}(\vec{x})e^{-i\omega_{\vec{k}_{\parallel,n}}x_{0}}+
a^{+}(\vec{k},n)\overset{{-}}{U}_{\vec{k}_{\parallel,n}}(\vec{x})e^{i\omega_{\vec{k}_{\parallel,n}}x_{0}} \Big)
\eqno{\rm{(1.12a)}}
$$
\vse

\noindent where

$$
U_{\vec{k}_{\parallel,n}}(\vec{x})=e^{i\vec{k}_{\parallel}\cdot\vec{x}_{\parallel}}U_{n}(z,D)
\eqno{\rm{(1.12b_{D})}}
$$
\vse

\noindent with

$$
U_{n}(z,D)=\sqrt{\frac{2}{d}\;}\sin\Big(\frac{n\pi}{d}z\Big)\;\;\;\;\;\;n=1,2,\dots
\eqno{\rm{(1.12c_{D})}}
$$
\vse

\noindent and $\vec{k}_{\parallel}\equiv(k_{x},k_{y})$, $\vec{x}_{\parallel}\equiv(x,y)$,

$$
\omega_{\vec{k}_{\parallel},n}\equiv \Bigg[ |\vec{k}_{\parallel}|^{2}+\Big(\frac{n\pi}{d}\Big)^2\Bigg]^{1/2}
\eqno{\rm{(1.12d_{D})}}
$$

for Dirichlet b.c., or
$$
U_{n}(z,P)=\sqrt{\frac{1}{d}\;}\exp(\frac{2in\pi}{d}z)
\eqno{\rm{(1.12b_{P})}}
$$
\vse
with
$$
\omega_{\vec{k}_{\parallel},n}\equiv \Bigg[ \vert\vec{k}_{\parallel}\vert^2+\Big(\frac{2n\pi}{d}\Big)^2\Bigg]^{1/2}\;n=\dots,-2,-1,0,1,2,\dots
\eqno{\rm{(1.12d_{P})}}
$$
\vse
for periodic b.c..
\noindent Above, $U_{\vec{k}_{\parallel,n}}$ are (improper) eigenfunctions of $(-\triangle)^{1/2}$, where
$\triangle$ denotes the Laplacean:

$$
(-\triangle)^{1/2}U_{\vec{k}_{\parallel,n}}(\vec{x})=\omega_{\vec{k}_{\parallel,n}}U_{\vec{k}_{\parallel,n}}
\eqno{\rm{(1.12e)}}
$$
\vse

\noindent The $a$, $a^{+}$ in (1.12a) are operator-valued distributions on $\mathcal{F}_{S}(\mathcal{H})$, where
$$
\mathcal{H}=L^2(\mathbb{R}^2)\otimes\mathcal{H}_{1}
\eqno{\rm{(1.13a)}}
$$
\vse

\noindent and
$$
\mathcal{H}_{1}=\Big \{ f \;|\; f(z)=l.i.m.\sum_{n=1}^{\infty}c_{n}U_{n}(z),
\textrm{ with } \sum _{n=1}^{\infty}|c_{n}|^2<\infty \Big \}
\eqno{\rm{(1.13b)}}
$$
\vse

\noindent where $l.i.m.$ denotes the limit in the topology of $L^2(0,d)$, and such that
$$
\big[ a(f),a^+(g) \big]= (f,g)_{\mathcal{H}} \mathbf{1}
\eqno{\rm{(1.14)}}
$$
\vse

\noindent on $\mathcal{F}_{0}$, where $(f,g)_{\mathcal{H}}$ denotes the scalar product on $\mathcal{H}$. The vacuum
$\Omega_{K_{d}}$ is defined by
$$
a(f)\Omega_{K_{d}}=0 \;\;\;\; \forall f \; \in \; \mathcal{H}
\eqno{\rm{(1.15)}}
$$
\vse
Analogous definitions apply in the case of periodic b.c..
The field $\Phi_{K_{d}}$ has the two-point function
$$
\Big[ \Phi_{K_{d,-}}(x), \Phi_{K_{d,+}}(x^{'})\Big]=\frac{1}{i}D_{K_{d}}^{(+)}(x,x^{'})
\eqno{\rm{(1.16)}}
$$
\vse

\noindent where $D_{K_{d}}^{(+)}$ is the distribution
$$
D_{K_{d}}^{(+)}(x,x^{'})=D_{K_{d}}^{(+)}(x_{0}-x_{0}^{'},\vec{x}, \vec{x}^{'})=\qquad\qquad\qquad\qquad\qquad\qquad\qquad
$$
$$
=i\sum_{n=1}^{\infty}\int\!\!\frac{d\vec{k}_{\parallel}}{2\omega_{\vec{k}_{\parallel,n}}}e^{i\vec{k}_{\parallel}\cdot
(\vec{x}_{\parallel}-\vec{x}_{\parallel}^{'})}U_{n}(z)U_{n}(z^{'})\!\!\!\!\!\!\!\!
\eqno{\rm{(1.17)}}
$$
\vse

Due to (1.16), (1.17), the canonical commutation relations (CCR) are altered with respect to their free field values
if $f,g \in \mathcal{S}(\mathbb{R}^{2})\otimes \mathcal{C}_{0}^{\infty}(0,d)\otimes\mathcal{S}(\mathbb{R})$
(corresponding to the variables $\vec{x}_{\parallel}$, $z$ and $x_{0}$), then

$$
\Big[ \Phi_{K_{d}}(f), \Phi_{K_{d}}(g)\Big]=0
\eqno{\rm{(1.18)}}
$$
\vse

\noindent not only when the points $x\equiv(\vec{x}_{\parallel}, z, x_{0}), x^{'}\equiv(\vec{x}_{\parallel}^{'}, z^{'}, x_{0}^{'})$
in the supports of $f$ and $g$ are space-like to one another, but also whenever $x$ is space-like to its mirror image
$(\vec{x}_{\parallel}^{'}, 2d-z^{'}, x_{0}^{'})$, i.e., $(\vec{x}_{\parallel}-\vec{x}_{\parallel}^{'})^{2}+(z+z^{'}-2d)^2-
(x_{0}-x_{0}^{'})^2>0$, corresponding to connecting the points $x$ and $x^{'}$ both by a ray and by one which suffers
a reflection on the right plate (there is, of course, an infinity of other possibilities involving the left plate and
multiple reflections). This is valid only for Dirichlet boundary conditions.

See also \cite{milloni} for the commutation relations for the electromagnetic field in the Coulomb gauge and Dirichlet b.c.,
and \cite{sommer} for a general discussion of boundary conditions in quantum field theory.

We need also regularized fields. Let $C:\mathbb{R}\longrightarrow\mathbb{R}$ be a smooth function (1.19a) satisfying

$$
C(0)=1
\eqno{\rm{(1.19b)}}
$$
$$
\int_{0}^{\infty}\!\!C^{(k)}(x)\,dx<\infty\;\;\;\;\;\forall \;k=1,2,\dots
\eqno{\rm{(1.19c)}}
$$
$$
\lim_{x \to \infty}C^{(k)}(x)=0\;\;\;\;\;\forall \;k=1,2,\dots
\eqno{\rm{(1.19d)}}
$$
\vse

We introduce a class of regularizing functions $C_{\Lambda}$ depending on a cutoff $\Lambda$ with dimensions of length by
$$
C_{\Lambda}(\vec{k})=C(\Lambda\omega_{\vec{k}})
\eqno{\rm{(1.20)}}
$$
\vse

\noindent where $\omega_{\vec{k}}$ is the frequency (1.8) (which in the case of $\Phi_{K_{d}}$, given by (1.12a),
is given by (1.12d)). By (1.19a) and (1.20),
$$
\lim_{\Lambda \to 0}C_{\Lambda}(\vec{k})=1
\eqno{\rm{(1.21)}}
$$
\vse

The regularized fields $\Phi_{\Lambda}^{(p)}$ and $\Phi_{K_{d},\Lambda}$ corresponding to the formal $\Phi^{(p)}$ and
$\Phi_{K_{d}}$ are defined by

$$
\Phi_{\Lambda}^{(p)}(x)=(2\pi)^{-p/2}\int\!\!\frac{d^{p}k}{\sqrt{2\omega_{\vec{k}}}}\Bigg(a_p(\vec{k})e^{-ik\cdot x}
\;\;+\;\;a_{p}^{+}(\vec{k})e^{ik\cdot x}\Bigg)\cdot C_{\Lambda}(\vec{k})
\eqno{\rm{(1.22)}}
$$

$$
\Phi_{K_{d},\Lambda}(x)=\frac{1}{(2\pi)^2}\int \!\! d\vec{k}_{\parallel} \sum_{n=1}^{\infty}\frac{1}
{\sqrt{2\omega_{\vec{k}_{\parallel,n}}}}\Bigg(a(\vec{k},n)U_{\vec{k}_{\parallel,n}}(\vec{x})
e^{-i\omega_{\vec{k}_{\parallel,n}}x_0}\;\;\;+\qquad\qquad\qquad\qquad
$$
$$
\qquad\qquad\qquad\qquad\qquad\qquad+\;\;\;a^{+}(\vec{k},n)\overset{{-}}{U}_{\vec{k}_{\parallel,n}}(\vec{x})e^{i\omega_{\vec{k}_{\parallel,n}}x_{0}} \Bigg)
\cdot C_{\Lambda}(\vec{k})
\eqno{\rm{(1.23)}}
$$
\vse

\noindent where $C_{\Lambda}$ is given by (1.19), (1.20), and has the property (1.21), as well as a normalization
condition inherited from (1.19b).

For the free field, we may, equivalently for (1.6), go from infinite space to a geometry with boundaries, by
expressing the normal ordering (1.4) in configuration space by the point splitting technique which yields

$$
:\Bigg(\frac{\partial\Phi}{\partial x_0}\Bigg)^2\!:\;\;\;=\;\;\;
\lim_{y \to x} : \frac{\partial \Phi(x)}{\partial x_{0}}\frac{\partial \Phi(y)}{\partial x_{0}}:\;\;\;=
\qquad\qquad\qquad\qquad\qquad\qquad
$$
$$
\qquad\qquad\qquad\qquad\qquad
=\;\;\;\lim_{y \to x}\frac{\partial \Phi(x)}{\partial x_{0}}\frac{\partial \Phi(y)}{\partial x_{0}}
\;\;\;+\;\;\;\frac{1}{i}\frac{\partial^{2}}{\partial x_{0}^{2}}D_{0,3}^{(+)}(x-y)
\eqno{\rm{(1.24)}}
$$
\vse

\noindent in the sense of operator-valued tempered distributions, where, in (1.24),

$$
\Phi \;\;\;\equiv\;\;\; \Phi^{(3)}
\eqno{\rm{(1.25)}}
$$

Finally, from (1.1) and (1.3),
$$
H(x)\;\;=\;\;\lim_{y \to x}\Bigg\{\frac{1}{2}\bigg[\frac{\partial\Phi(x)}{\partial x_{0}}\frac{\partial\Phi(y)}{\partial y_{0}}\;\;+\;\;
\nabla\Phi(x)\cdot\nabla\Phi(y)\bigg]\;\;+
\qquad\qquad\qquad\qquad\qquad
$$
$$
\qquad\qquad\qquad\qquad\qquad\qquad
+\;\;\frac{1}{i}\frac{\partial^{2}}{\partial x_{0}^{2}}\;\;D_{0,3}^{(+)}(x-y)\Bigg\}
\eqno{\rm{(1.26)}}
$$
\vse

\noindent which we take to be the Hamiltonian density (operator) describing the field, both free and with boundaries,
in agreement with (1.5) and the discussion following it. In case we wish to describe the field with boundaries, the first
three terms in (1.26) must be defined on symmetric Fock space on the adequate (one-particle) Hilbert
space $\mathcal{H}$, i.e., the concrete representation of the field operator is dictated by the geometry. In the case
of parallel plates, i.e., with the fields defined on $K_{d}$, given by (1.2), with Dirichlet b.c. on $\partial K_{d}$,
given by (1.3), the field is (formally) given by (1.12), on $\mathcal{F}_{S}(\mathcal{H})$, given by (1.13).

Thus, (1.26) has to be an operator on $\mathcal{F}_{S}(\mathcal{H})$, and therefore must be represented in the form:

$$
H_{K_{d}}\;;=\;\frac{1}{2}\;\; ;\bigg(\frac{\partial\Phi(x)}{\partial x_{0}}\bigg)^2;\;\;+\;\;\frac{1}{2}\;\;
;\nabla\Phi(x)\cdot\nabla\Phi(x);\;\;+
\qquad\qquad\qquad\qquad\qquad
$$
$$
\qquad\qquad\qquad\qquad\qquad\qquad\qquad
+\;\; \frac{1}{i}\;\lim_{y \to x}\;\frac{\partial^2}{\partial x_{0}^{2}}\bigg\{D_{0,3}^{(+)}(x-y)-D_{K_{d}}^{(+)}(x-y)\bigg\}
\eqno{\rm{(1.27)}}
$$
\vse

\noindent where $D_{K_{d}}^{(+)}$ is given by (1.17) and the semicolons in (1.27) denote normal ordering, with respect
to the emission and absorption operators in (1.12a) (which satisfy (1.14)) and the vacuum $\Omega_{K_{d}}$, defined
by (1.15).

Taking into account (1.22), (1.23), we define the quantities corresponding to (1.1), (1.10) (in $p$ dimensions), (1.27) and
(1.17), denoting them by $h^{(p)}(x), D_{0,\Lambda}^{(+)}, H_{\Lambda, K_{d}}(x)$ and $D_{K_{d}}^{(+)}$, respectively.
$H_{\Lambda,K_{d}}$ represents the regularized Hamiltonian density in a theory with boundary $\partial K_{d}$ given
by (1.3).

Let, now, $p=2$, and $\mathcal{H}_r$ $(\textrm{resp. }\mathcal{H}_l)$ denote two copies of $L^2(\mathbb{R}^2)$
corresponding to the right (resp. left) plates (components of $\partial K_{d}$), $\mathcal{F}_S(\mathcal{H}_r)$ and
$\mathcal{F}_S(\mathcal{H}_l)$ the related symmetric Fock spaces, with vacua $\Omega_r$, $\Omega_s$, satisfying (1.11b)
(with $p=2$). The related Hamiltonian densities will be denoted by $h_{\Lambda,r}^{(2)}(x)$ and $h_{\Lambda, l}^{(2)}(x)$. They will make their appearance only in section 4.

Further, let $\delta_r^{(2)}$, $\delta_l^{(2)}$ denote the delta-functions associated to the right and left
plates (see \cite{guelfand}, Ch.3, $\S$ 1, for delta-distributions and other singular functions associated to a regular
surface).

Finally, let $E_{\Lambda}^{vac}$ denote the vacuum energy density corresponding to $H_{\Lambda,K_{d}}$. By (1.27), it is
(with $\Omega_{K_{d}}$ defined by (1.15)):

$$
E_{\Lambda}^{vac}(x)\;\;=\;\;(\Omega_{K_{d}}, H_{\Lambda,K_{d}}(x)\;\Omega_{K_{d}})\;\;=
\qquad\qquad\qquad\qquad\qquad\qquad
$$
$$
\qquad\qquad\qquad\qquad
=\;\;\frac{1}{i}\lim_{y \to x}\Bigg\{\frac{\partial^2}{\partial x_0^2}\bigg[D_{0,3,\Lambda}^{(+)}(x-y)\;\;-
\;\;D_{K_{d},\Lambda}^{(+)}(x,y)\bigg]\Bigg\}
\eqno{\rm{(1.28)}}
$$
\vse

\noindent {\bf 2. \ Poisson's summation formula and the Casimir energy density}

In this section we use for convenience a special regularizer,    

$$
C(x)\;\;=\;\;e^{-x}\;\;\;\;\;x\ge0, x\in\mathbb{R}
\eqno{\rm{(2.1)}}
$$

which satisfies the conditions (1.19a), (1.19b) and (1.19c). In the
next section we shall demonstrate that there is no loss in choosing
any regularizer defining the above conditions for the calculation of
the Casimir energy density. Putting (2.1) into (1.28) we obtain

$$
E_{\Lambda}^{vac}(\vec{x})\;=\;\frac{1}{2}\frac{\partial}{\partial\Lambda}\Bigg\{\frac{1}{(2\pi)^3}\int\!\!d^3k\;
e^{-i\big[k_0\tau-\vec{k}\cdot(\vec{x}-\vec{y})\big]_{\vec{y}=\vec{x}}^{\tau=0}}\times e^{-\Lambda k_{0}}\;-
\qquad\qquad\qquad
$$
$$
\qquad\qquad\qquad\qquad\qquad
-\;\int\!\!dk_1\;\int\!\!dk_2\;\sum_{n}|U_{n}(\vec{x})|^2\times\;
e^{-\Lambda \omega_n}\Bigg\}
\eqno{\rm{(2.2)}}
$$
\vse

\noindent with, for the parallel plates, $U_n(\vec{x})$ either of 
$$
U_{n,D}(\vec{x})\;=\;\frac{1}{2\pi}\sqrt{\frac{2}{d}\;}\sin\big(\frac{n\pi}{d}z\big)e^{i(k_xx+k_yy)}\;n=1,2,\dots
\eqno{\rm{(2.3_{D})}}
$$
\vse
or
$$
U_{n,P}(\vec{x})\;=\frac{1}{2\pi}\sqrt{\frac{1}{d}}\exp\big(\frac{2in\pi}{d}z\big)e^{i(k_xx+k_yy)}\; n=\dots,-2,-1,0,1,2,\dots
\eqno{\rm{(2.3_{P})}}
$$
\noindent where $\;\;\omega_n^2=k_1^2+k_2^2+\big(\frac{n\pi}{d}\big)^2\;$ for Dirichlet b.c. and  $\;\;\omega_n^2=k_1^2+k_2^2+\big(\frac{2n\pi}{d}\big)^2\;$ for periodic b.c., using now and henceforth the letters D and P to denote Dirichlet and periodic b.c.. We first consider Dirichlet b.c.. From (2.2) we may write

$$
E_{\Lambda}^{vac}(\vec{x})\;=\;\frac{1}{2}\frac{\partial}{\partial\Lambda}\Bigg\{-\frac{1}{(2\pi)^3}\int\!\!d^3k\;e^{-\Lambda k}\;+\;
I(\Lambda)\Bigg\}
\eqno{\rm{(2.4)}}
$$
\vse
where $k=\vert\vec{k}\vert$, and

$$
I(\Lambda)\equiv\frac{1}{(2\pi)^2}\,\frac{2}{d}\sum_{n=0}^{\infty}\int\!\!dk_1\int\!\!dk_2\;
sin^2\big(\frac{n\pi}{d}z\big)e^{-\Lambda \omega_n}\;=
\qquad\qquad\qquad\qquad\qquad\qquad
$$
$$
\qquad\qquad\qquad\qquad\qquad\qquad
=\;\frac{1}{(2\pi)^2d}\sum_{n=-\infty}^{+\infty}\int\!\!dk_1\int\!\!dk_2\;
sin^2\big(\frac{n\pi}{d}z\big)e^{-\Lambda \omega_n}
\eqno{\rm{(2.5)}}
$$
\vse
The summation term in the energy density in (2.4) can be extended to accommodate negative terms in $n$, noticing that
the parity of integrand is even and the term $n=0$ yields zero. Let now, by Poisson's summation formula (PSF) (see, e.g., \cite{graham}),
$$
\sum_{n=-\infty}^{+\infty}\!\!f(2\pi n)\;=\;\frac{1}{\sqrt{2\pi\;}}\;\sum_{m=-\infty}^{+\infty}\!\!\hat{f}(m)
\eqno{\rm{(2.6a)}}
$$
\noindent where
$$
\hat{f}(m)\;=\;\frac{1}{\sqrt{2\pi\;}}\int_{-\infty}^{+\infty}\!\!\!\!\!dk\,e^{-imk}f(k)\;\textrm{.}
\eqno{\rm{(2.6b)}}
$$
\noindent and
$$
f(2\pi n)\equiv sin^2\big(\frac{n\pi}{d}z\big)e^{-\Lambda\omega_n}
\eqno{\rm{(2.6c)}}
$$
\vse

\noindent we obtain from (2.5)
$$
I\;\equiv\frac{1}{2}\frac{\partial I(\Lambda)}{\partial \Lambda}=\frac{1}{2(2\pi)^3d}\sum_{m=-\infty}^{\infty}\int_{-\infty}^{+\infty}\!\!\!\!\!\!dk_1
\int_{-\infty}^{+\infty}\!\!\!\!\!\!dk_2\int_{-\infty}^{+\infty}\!\!\!\!\!\!dk_3^{'}
\sqrt{k_1^2+k_2^2+\Big(\frac{k_3^{'}}{2d}\Big)^{2}}\;\;\times
\qquad\qquad\qquad\qquad
$$
$$
\qquad\qquad\qquad\qquad
\times\;\; sin^2\Big(\frac{k_3^{'}}{2d}z\Big)
e^{-\Lambda\sqrt{k_1^2+k_2^2+\Big(\frac{k_3^{'}}{2d}\Big)^2}}e^{-imk_3^{'}}
\eqno{\rm{(2.7)}}
$$
\vse

Performing a change of variable $k_3=k_3^{'}/2d$ on (2.7) we get
$$
I=I_1+I_2
\eqno{\rm{(2.8)}}
$$
with
$$
I_1\;=\;\frac{1}{2(2\pi)^3}\sum_{m=-\infty}^{+\infty}\int_{-\infty}^{+\infty}\!\!\!\!\!\!dk_1
\int_{-\infty}^{+\infty}\!\!\!\!\!\!dk_2\int_{-\infty}^{+\infty}\!\!\!\!\!\!dk_3\;
\sqrt{k_1^2+k_2^2+k_3^2}e^{-\Lambda\sqrt{k_1^2+k_2^2+k_3^2}}e^{-i2mdk_3}
\eqno{\rm{(2.9)}}
$$
$$
I_2\;=\;-\frac{1}{2(2\pi)^3}\sum_{m=-\infty}^{+\infty}\int_{-\infty}^{+\infty}\!\!\!\!\!\!dk_1
\int_{-\infty}^{+\infty}\!\!\!\!\!\!dk_2\int_{-\infty}^{+\infty}\!\!\!\!\!\!dk_3\;
\sqrt{k_1^2+k_2^2+k_3^2}\;\times
\qquad\qquad\qquad\qquad\qquad
$$
$$
\qquad\qquad\qquad\qquad\qquad\qquad
\times \; cos(2k_3z)e^{-\Lambda\sqrt{k_1^2+k_2^2+k_3^2}}e^{-i2mdk_3}
\eqno{\rm{(2.10)}}
$$
\vse

By (2.4), (2.8), (2.9) and (2.10) we obtain
$$
E_{\Lambda}^{vac}(z,d;D)\;=\;\frac{1}{2}\Bigg\{\frac{4}{(2\pi)^2}\frac{(2z)^2-3\Lambda^2}{[\Lambda^2+(2z)^2]^3}\;-\;
\frac{4}{(2\pi)^2}\sum_{m\ne0}\frac{(2md)^2-3\Lambda^2}{[\Lambda^2+(2md)^2]^3}\;+\;
$$
$$
+\;\frac{2}{(2\pi)^2}\sum_{m\ne0}\Bigg[\frac{\big[2(md+z)\big]^2-3\Lambda^2}{\big[\Lambda^2+[2(md+z)]^2\big]^3}\;+\;
\frac{\big[2(md-z)\big]^2-3\Lambda^2}{\big[\Lambda^2+[2(md-z)]^2\big]^3}\Bigg]\Bigg\}
\eqno{\rm{(2.11)}}
$$
\vse

\noindent where the first term above is the term with $m=0$ in $I_2$. The $m=0$ term in $I_1$ cancels exactly the
first integral in (2.4). By (2.11),

$$
E^{vac}(z,d;D)\equiv \lim_{\Lambda \to 0^+}E_{\Lambda}^{vac}(z,d)\;=
\;\frac{1}{2}\Bigg\{\frac{1}{4(2\pi)^2}\frac{1}{z^4}-
\qquad\qquad\qquad\qquad\qquad\qquad
$$
$$
\qquad\qquad\qquad\qquad
-\frac{1}{2(2\pi)^2}\sum_{m=1}^{\infty}\frac{1}{(md)^4}
+\frac{1}{4(2\pi)^2}\sum_{m=1}^{\infty}\bigg[\frac{1}{(md+z)^4}+
\frac{1}{(md-z)^4}\bigg]\Bigg\}
\eqno{\rm{(2.12a)}}
$$
The first term in (2.12a) diverges on the plate where $z=0$,
the last term (for $m=1$) diverges on the other plate where $z=d$. These dominant terms
can be identified as stationary points in the PSF by (2.7):

Indeed, writing $sin^2\big(\frac{k^{'}_3}{2d}z\big)=\Big(\frac{e^{i\frac{k_3^{'}}{2d}z}-
e^{-i\frac{k_3^{'}}{2d}z}}{2i}\Big)^2$ in (2.7), these points correspond to $\frac{\partial}{\partial k_3^{'}}
\big[(md\pm z)k_3^{'}\big]=0$, which lead to $z=0$ (for $m=0$) and $z=d$ (for $m=\pm 1$).

A general method combining the PSF with the stationary phase method for obtaining asymptotic estimates was developed in
\cite{graham} (see also \cite{berry}).

We now consider periodic b.c.. in this case it is readily seen from $(2.3_{P})$ that $I(\Lambda)$ in (2.5) is given by the same expression on the r.h.s. of (2.5) with, however, $\sin(\frac{n\pi}{d}z)^2$ replaced by one. This implies that $I_{2})=0$ in (2.8), and $I_{1}$ is given by the same expression on the r.h.s. of (2.9), but with $\exp(-2imdk_3)$ replaced by $\exp(-imdk_3)$. The final result is
\vse
$$
E^{vac}(z,d;P)\equiv \lim_{\Lambda \to 0^+}E_{\Lambda}^{vac}(z,d;P)\;
-\frac{1}{\pi^2d^4}\zeta(4)=-\frac{\pi^2}{90d^4}
\eqno{\rm{(2.12b)}}
$$
\vse
Thus,(2.12) is the result provided by lqft for the original problem. We define 
$$
E^{vac}(z,d) = 0   \mbox{ for } z<0  \mbox{ or } z>d
\eqno{\rm{(2.13)}}
$$
corresponding to the fact that the outer Casimir problem yields zero contribution to the energy density and pressure ( Remark 3.7 ).

Consider, now, a compact subset $K_{d}^{A}$ of $K_{d}$ (defined by (1.2)):
$$
K_{d}^{A}\equiv\bigg\{\vec{x}\equiv(\vec{x}_{\parallel},z)\textrm{, with }\vec{x}_{\parallel}(x,y)\in
\big[-\frac{L_1}{2},\frac{L_1}{2}\big]\times\big[-\frac{L_2}{2},\frac{L_2}{2}\big]\textrm{ and }
0\le z \le d \bigg\}
\eqno{\rm{(2.14)}}
$$
\vse

\noindent with
$$
L_1L_2\;=\;A
\eqno{\rm{(2.15)}}
$$
\vse

Let $\big\{\chi_{K_{d}^{A}}^{(n)}(.)\big\}_{n=1,2,\dots}$ be a sequence of smooth functions approaching, as $n\to\infty$,
the characteristic function of $K_{d}^{A}$, and such that
$$
\chi_{K_{d}^{A}}^{(n)}(\vec{x})\;=\;1\;\;\;\textrm{ for }\;\vec{x}\;\in\;K_{d}^{A,O}\;\cup\;\partial K_{d}^{A}
\;\;\;\forall n=1,2,\dots
\eqno{\rm{(2.16)}}
$$
\vse

\noindent where $K_{d}^{A,O}$ denotes the interior of $K_{d}^{A}$.

We define the vacuum energy $E^{vac}(A)$ of the region $K_{d}^{A}$ by

$$
\lim_{\Lambda \to 0+} \;\; \lim_{n\to\infty}\int \!\! d^{3}x \;\chi_{K_{d}^{A}}^{(n)}(\vec{x})E_{ren,\Lambda}^{vac}(x_0,\vec{x})
\equiv E^{vac}(A,d)
\eqno{\rm{(2.17)}}
$$
\vse

\noindent whenever the double-limit on the l.h.s. of (2.17) exists, is independent of the regularization (1.19), (1.20)
and does not depend on $x_0$. The \underline{vacuum energy per unit area} $\varepsilon$ -- a basic observable
quantity -- is, then, defined by
$$
\varepsilon\;\;\equiv\;\;\lim_{A\to\infty}\;\frac{E^{vac}(A,d)}{A}
\eqno{\rm{(2.18)}}
$$
\vse

\noindent whenever the limit on the r.h.s. of (2.18) exists. By (2.12a), $E^{vac}(z,d;D)$ is not integrable over $0\le z\le d$, and therefore
$$
\varepsilon_D = -\infty
\eqno{\rm{(2.19a)}}
$$
\vse
However, by (2.12b), for periodic b.c.,
$$
\varepsilon_P = -\frac{\pi^2}{90d^3}
\eqno{\rm{(2.19b)}}
$$
\vse
The above differs from the finite part (3.17b) of $\varepsilon_D$ by a factor 
$2^{-4}$, due to the mentioned replacement of the summand $2m$ in $\exp(-2imdk_3)$ by $\exp(-1mdk_3)$ in (2.9).

(2.19a) means that lqft rejects Dirichlet (Neumann, mixed) b.c. in the Casimir effect for a scalar field (see also the last paragraph of section 5). For the moment, we just remark that this is consistent with the fact that
formula (1.6) has being arrived at by the assumptions of local quantum theory (\cite{kay}, appendix), in particular
the assumption that the states $S$ and $\omega$ are locally quasi-equivalent (\cite{haag}, pg.131). This is
not true everywhere, there are singular points or surfaces (in our case in $\mathbb{R}^3$) in the neighborhood
of which even such local quasi-equivalence must fail to hold, otherwise the renormalized one-point function (1.6b) $\omega_{1,+}(x,x) = \omega_{+,1}(z)$ which depends only on z (the two equal time variables being arbitrary), would exist as a Schwartz distribution by the ingenious discussion in the Appendix of \cite{kay}, which we have seen \textbf{not} to be true by (2.12a) in the case of Dirichlet b.c. (a similar proof holds for Neumann or mixed b.c). Local quasiequivalence is essential to thermodynamics and of great importance in the context of quantum field theory in curved space-time (which means that it should also play an important role in cosmology), see the discussion and references in \cite{haag}. 

In the next section 3 we provide an alternate picture of why Dirichlet and periodic b.c. yield results which are so qualitatively different in the context of cutoff (regularization) independence of $\varepsilon$, where we show the emergence of the RDST mentioned in the introduction for Dirichlet but \textbf{not} periodic b.c. (remark 3.6). Notice that the forthcoming (3.17a,b) are consistent with (2.19a)! Section 4 relates these surface terms to Hadamard regularization of (2.12a), in agreement with previous work on the subject, also mentioned in the introduction. We leave to the last three paragraphs of section 5 (conclusion) a brief but important discussion of why we believe that the message of lqft, our starting point, is clear and correct also in the context of the Casimir effect.

\noindent {\bf 3. \ The cutoff independence of the energy per unit area}

\vspace{0.3cm}

We now write (1.28), with a general cutoff satisfying (1.19) and (1.20), as

$$
E_{\Lambda}^{vac}\;=\;\frac{A}{2(2\pi)^2}\Bigg\{-2d\int_{0}^{\infty}\!\!\!dk\;k^3C(\Lambda k)\;+
\qquad\qquad\qquad\qquad\qquad\qquad\qquad
$$
$$
\qquad\qquad\qquad
+\;2\pi\sum_{n=1}^{\infty}\int_{0}^{\infty}\!\!\!dk\;k
\sqrt{\Big(\frac{n\pi}{d}\Big)^2+k^2\;}C\Bigg(\Lambda\sqrt{\Big(\frac{n\pi}{d}\Big)^2+k^2\;}\Bigg)\Bigg\}
\eqno{\rm{(3.1)}}
$$
\vse

Rewriting the second term of the r.h.s. in (3.1) with a change of variable we have

$$
E_{\Lambda}^{vac}\;=A\Bigg\{-\frac{d}{(2\pi)^2}\int_{0}^{\infty}\!\!\!dk\;k^3C(\Lambda k)\;
+\;\frac{1}{8\pi}\lim_{n \to \infty}\;\sum_{m=1}^{n}g(m)\Bigg\}
\eqno{\rm{(3.2)}}
$$
\vse

\noindent with
$$
g(m)\;=\;\int_{0}^{\infty}\!\!\!du\sqrt{u+\Big(\frac{m\pi}{d}\Big)^2}\;C\Bigg(\Lambda\sqrt{u+\Big(\frac{m\pi}{d}\Big)^2}\Bigg)\;=
$$
$$
=\;\int_{(m\pi/d)^2}^{\infty}\!\!\!\!du\sqrt{u}C(\Lambda\sqrt{u})
\qquad\qquad\qquad\;\;\;\;\;
\eqno{\rm{(3.3)}}
$$
\vse

Now, we can introduce the Euler-Maclaurin sum formula, which yields (\cite{hardy}, p.326) under assumptions
(1.19c) and (1.19d):

$$
\sum_{m=1}^{n}g(m)\;-\;\frac{2d}{\pi}\int_{0}^{\infty}\!\!\!dq\;q^3C(\Lambda q)-\frac{1}{2}g(0)\to\Sigma_k
\eqno{\rm{(3.4)}}
$$
\vse

\noindent where
$$
\Sigma_k\;=\;-\;S_k(0)\;-\;\frac{1}{(2k+2)!}\int_{0}^{\infty}\!\!\!dt\;\psi_{2k+2}(t)g^{(2k+2)}(t)
\textrm{, }\;\;\;k=1,2,\dots
\eqno{\rm{(3.5)}}
$$

\noindent with
$$
S_k(0)\;=\;\sum_{r=1}^{k}(-1)^{r-1}\frac{B_r}{(2r)!}g^{(2r-1)}(0)\textrm{,}
\eqno{\rm{(3.6)}}
$$

\noindent and

$$
\psi_k(t)\;=\;\phi_k(t)\textrm{ mod } 1
\eqno{\rm{(3.7)}}
$$

\noindent where $\phi_k$ can be obtained as follows
$$
x\frac{e^{xt}-1}{e^x-1}\;=\;\sum_{n=1}^{\infty}\phi_k(t)\frac{x^n}{n!}\textrm{.}
\eqno{\rm{(3.8)}}
$$
\vse

\underline{Theorem 3.1}

Under assumptions (1.19c) and (1.19d), $\Sigma_k$ (with $k=1,2,\dots$) is independent of $k$, i.e.
$$
\Sigma_k\;=\;\Sigma_{k+1}\qquad\forall\;k=1,2,\dots
\eqno{\rm{(3.9)}}
$$
\vse

\underline{Proof}. By (3.5)
$$
\Sigma_{k+1}\;=\;-S_{k+1}(0)-\frac{1}{(2k+4)!}\int_{0}^{\infty}\nsp
dt\;\psi_{2k+4}(t)g^{(2k+4)}(t)
\eqno{\rm{(3.10)}}
$$
\vse

The first term on the r.h.s. of (3.10) may be written
$$
S_{k+1}(0)\;=\;\sum_{r=1}^{k+1}(-1)^{r-1}\frac{B_r}{(2r)!}g^{(2r-1)}(0)\;=
\qquad\qquad\qquad\qquad\qquad\qquad\qquad\qquad\qquad
$$
$$
=\;\sum_{r=1}^{k}(-1)^{r-1}\frac{B_r}{(2r)!}g^{(2r-1)}(0)\;+\;(-1)^k\frac{B_{k+1}}{[2(k+1)]!}g^{[2(k+1)-1]}(0)\;=
$$
$$
=\;S_k(0)+(-1)^k\frac{B_{k+1}}{(2k+2)!}g^{(2k+1)}(0)\textrm{.}
\qquad\qquad\qquad\qquad\qquad\qquad\;
\eqno{\rm{(3.11)}}
$$
\vse

By (3.7) and (3.8) it follows that (see \cite{hardy}, pp.320-321)

$$
\psi_k(0)\;=\;0\textrm{,}
\eqno{\rm{(3.12a)}}
$$
$$
\psi_{2m-1}(x)=\frac{\psi_{2m}^{'}(x)}{2m}\;\textrm{,}
\eqno{\rm{(3.12b)}}
$$
\noindent and
$$
\psi_2m(x)=\frac{\psi_{2m+1}^{'}(x)}{2m+1}+(-1)^mB_m\;\textrm{.}
\eqno{\rm{(3.12c)}}
$$
\vse

Rewriting the second term on the r.h.s. in (3.5) using (3.12c) with $m=k+1$ and integrating by parts using (1.19d),
we have:

$$
\Sigma_{k+1}\;+\;S_{k+1}(0)\;=
\;\frac{1}{(2k+2)!}\Bigg[-\frac{1}{2k+3}\int_{0}^{\infty}\nsp dt\;\psi_{2k+3}(t)g^{(2k+3)}(t)\;+
\qquad\qquad\qquad\
$$
$$
\qquad\qquad\qquad\qquad\qquad\qquad\qquad\qquad
+\;(-1)^kB_{k+1}g^{(2k+1)}(0)\Bigg]
\eqno{\rm{(3.13)}}
$$
\vse

Using (3.12b) above, with $m=k+2$, and integrating by parts again we have

$$
\Sigma_{k+1}\;+\;S_{k+1}(0)\;=\;
\frac{1}{(2k+4)!}\int_{0}^{\infty}\nsp dt\;\psi_{2k+4}(t)g^{(2k+4)}(t)\;+\;\frac{(-1)^kB_{k+1}g^{(2k+1)}(0)}{(2k+2)!}
\eqno{\rm{(3.14)}}
$$
\vse

\noindent which yields
$$
\Sigma_{k+1}\;+\;S_{k+1}(0)\;=
\frac{1}{(2k+4)!}\int_{0}^{\infty}\nsp dt\;\psi_{2k+4}(t)g^{(2k+4)}(t)\;=
\qquad\qquad\qquad\qquad\qquad
$$
$$
\qquad\qquad\qquad\qquad
=\;\frac{1}{(2k+2)!}\int_{0}^{\infty}\!\!\!dt\;\psi_{2k+2}(t)g^{(2k+2)}(t)\;-\;
\frac{(-1)^kB_{k+1}g^{(2k+1)}(0)}{(2k+2)!}
\eqno{\rm{(3.15)}}
$$
\vse

Finally, putting (3.15) and (3.11) into (3.10), we find
$$
\Sigma_{k+1}\;=\;-S_k(0)\;-\;\frac{1}{(2k+2)!}\int_{0}^{\infty}\nsp dt\;\psi_{2k+2}(t)g^{(2k+2)}\;=
$$
$$
=\;\Sigma_k\;\textrm{.}
\qquad\qquad\qquad\qquad\qquad\qquad\qquad\;\;\;\;\;\,
\eqno{\rm{(3.16)}}
$$
\vse

\underline{Theorem 3.2}

Let in addition to (1.19c) and (1.19d), the normalization condition (1.19b) hold. Then 
$$
\frac{E_{\Lambda}^{vac}}{A}\;=\;-\frac{1}{8\pi\Lambda^3}\int_0^{\infty}\!\!du\;u^2C(u)\;+\varepsilon_{f}\;+\;O(\Lambda^2)
\eqno{\rm{(3.17a)}}
$$
where
$$
\;\;\;\;\;\varepsilon_{f}\;\;=\;\;-\frac{1}{2}\frac{\pi^2}{720d^3}
\eqno{\rm{(3.17b)}}
$$
with $\varepsilon_{f}$ independent of the regularizer, if the latter satisfies (1.19).

\underline{Proof}. By (3.2), (3.3), (3.4) and (3.16):

$$
\frac{E_{\Lambda}^{vac}}{A}\;=\;\lim_{n\to\infty}\;\frac{1}{8\pi}\sum_{m=1}^{\infty}g(m)\;=
\;\frac{1}{8\pi}\bigg[\frac{1}{2}g(0)\;+\;\Sigma_2\bigg]
\eqno{\rm{(3.17c)}}
$$
\vse

\noindent where, by (3.5),
$$
\Sigma_2\;=\;-\frac{B_1}{2}g^{(0)}(0)\;+\;\frac{B_2}{24}g^{(3)}(0)\;-
\;\frac{1}{6!}\int_0^{\infty}\!\!\psi_6(t)g^{(6)}(t)dt
\eqno{\rm{(3.18)}}
$$
\vse

By (3.3),
$$
\frac{1}{2}g(0)\;=\;\frac{1}{2}\int_0^{\infty}\!\!du\;\sqrt{u}C(\Lambda\sqrt{u})\;=
\;\frac{1}{\Lambda^3}\int_0^{\infty}\!\!du\;u^2C(u)
\eqno{\rm{(3.19)}}
$$
\vse

Again, by (3.3),
$$
\frac{d}{\pi}g^{(1)}(m)\;=\;-2\Big(\frac{m\pi}{d}\Big)^2C\Big(\Lambda\frac{m\pi}{d}\Big)
\eqno{\rm{(3.21a)}}
$$

$$
\frac{d}{\pi}g^{(3)}(m)\;=\;-4\Big(\frac{\pi}{d}\Big)^2C\Big(\Lambda\frac{m\pi}{d}\Big)\;-
\qquad\qquad\qquad\qquad\qquad\qquad\qquad\qquad\qquad
$$
$$
\qquad\qquad\qquad\qquad
-\;8\frac{\pi}{d}\Big(\frac{m\pi}{d}\Big)\Big(\frac{\Lambda\pi}{d}\Big)C^{(1)}\Big(\Lambda\frac{m\pi}{d}\Big)-
2\Big(\frac{m\pi}{d}\Big)^2\Big(\frac{\Lambda\pi}{d}\Big)^2C^{(2)}\Big(\Lambda\frac{m\pi}{d}\Big)
\eqno{\rm{(3.21b)}}
$$
\vse

By (3.18), (3.19), (3.21) and (1.19b),
$$
\Sigma_2\;=\;-\;\frac{B_2}{6}\Big(\frac{\pi}{d}\Big)^3\;+\;O(\Lambda^2)
\eqno{\rm{(3.22)}}
$$
 
Putting (3.19) and (3.22) into (3.17c) we obtain ,using $B_2=\frac{1}{30}$, (3.17a), with $\varepsilon_{f} $ given by (3.17b).

Remark 3.1}: It is remarkable that in (3.4) the second term in the l.h.s. is the vacuum term, and the third
one, the surface term, appearing in a natural way as necessary subtractions in a purely mathematical context. This surface term will be reconsidered in the next section.

\underline{Remark 3.2}: Usually the result is presented informally without the important last term in (3.5), and assuming that $C$ satisfies
$C^{(k)}(0)=0$ for all $k\ge1$ besides (1.19b), which is not satisfied by the special and important choice (2.1)
(\cite{itzykson}, p.138). See, however, ref. \cite{plunien} for a much nicer approach to the subject.

\underline{Remark 3.3}: $\Sigma_k$ ($k\ge1$) is called the ($\mathfrak{R},0$) sum of the (divergent) series $\sum_{m=1}^{\infty}g(m)$,
where $\mathfrak{R}$ refers to Ramanujan and $0$ to the reference point (the origin in our case).

\underline{Remark 3.4}: Other classes of regularizers do not necessarily lead to regularization independence. See \cite{kay} for
thorough discussion of several types of regularization. We believe the present class is ``natural'' from the physical
point of view, because it simulates a dielectric constant with suitable behavior at high energies.

\underline{Remark 3.5}: The idea of the proof of Theorem 4.2 appeared in the second reference under \cite{manzoni}, Theorem p.319, but only an
incomplete sketch was given there. More importantly, however, the term $\frac{1}{2}g(0)$ was (wrongly) asserted there
to contribute only to the $\Lambda$-independent terms, while it is precisely this term that yields the surface
contribution.

\underline{Remark 3.6}: The surface term in (3.17a) is absent for periodic b.c., because the latter allows for the $m=0$
term in (3.17c), which cancels it exactly. This explains the result of \cite{scharf}, which bears
some similarity with the also very special model in \cite{kay}.

\underline{Remark 3.7}: The \underline{external} Casimir energy is zero, see \cite{manzoni}, whose proof remains unaltered.
Result (3.17a) is one-half of the result for the electromagnetic field, due to summation over the two polarization
states in the latter. Polarization does, however, play a major role in explaining the cancelation occurring
in \cite{kuhn}.

\underline{Remark 3.8}: The series (3.17a) is a \underline{divergent} asymptotic series, but the rest in (2.13) is -- by (3.17c) -- bounded by constant $\Lambda^2$, so that its limit as $\Lambda\to0+$
exists and is zero.

\underline{Remark 3.9}: The present approach also works for the inner problem for the cube, see \cite{wreszinski}

\vspace{0.7cm}

\noindent {\bf 4. \ Symanzik counterterms versus Hadamard regularization}

In the previous section we saw that RI of the energy per unit area is not true in the Dirichlet case unless a regularization-dependent surface term is subtracted from the (regularized) energy density: the second term on the r.h.s. of (3.17a)! This term corresponds exactly to the replacement of the regularized vacuum energy density (2.2) by
$$
E_{ren,\Lambda}^{vac}\;\;\equiv\;\;E_{\Lambda}^{vac}(x)\;\;+\;\;\frac{1}{4}\Bigg[\bigg(\Omega_{r},h_{\Lambda,r}^{(2)}
\;\Omega_{r}\bigg)\;\delta_{r}^{(2)}(\vec{x})\;+\;\bigg(\Omega_{l},h_{\Lambda,l}^{(2)}\;\Omega{l}\bigg)\delta_{l}^{(2)}
(\vec{x})\Bigg]
\eqno{\rm{(4.1)}}
$$
\vse
The additional terms in (4.1) may be called ``surface renormalization counterterms''. they have been introduced by Symanzik \cite{symanzik} in a different framework of quantum field theory in the Schroedinger picture and including interactions. In (4.1), the Hamiltonian densities  $h_{\Lambda,r}^{(2)}(x)$ and $h_{\Lambda, l}^{(2)}(x)$ and the vacua $\Omega_r$, $\Omega_s$ have been defined in section 1, and correspond to the Hamiltonian densities and vacua of regularized quantum fields in (infinite) two-dimensional space ($p=2$ in (1.22)). Thus, in (4.1),
$$
(\Omega_r, h_{\Lambda,r}^{(2)}\;\Omega_r)\;=\;\frac{1}{2}\frac{2\pi}{(2\pi)^2}\int_{0}^{\infty}\!\!dk\;k^2 C(\Lambda k)
\eqno{\rm{(4.2)}}
$$
\vse
and similarly for the left index. Finally,
$$
E_{ren,\Lambda}^{vac}\;\equiv\;\lim_{n\to\infty}\int\!\!d^3x\chi_{K_{d}^{A}}^{(n)}(\vec{x})
E_{ren,\Lambda}^{vac}(x_0,\vec{x})
\eqno{\rm{(4.3)}}
$$
\vse
By (4.1), (4.3), (3.17a,b),
$$
E_{ren,\Lambda}^{vac}\;=\;-\frac{1}{2}\frac{\pi^2A}{720d^3}+O(\Lambda^2)
\eqno{\rm{(4.4)}}
$$
\vse

By (4.1) and (4.2) we see that the Symanzik term is a surface term, which diverges as the cutoff is removed. However, it is there and, as remarked by \cite{jaffe}, cannot be disentangled from the finite part (4.4).

However, we have a surprise: the same value of $\varepsilon_{f}$ ((3.17b) or (4.4)) is obtained by Hadamard regularization of the divergent integral of $ E^{vac}(z,d;D)$ given by (2.12a) - the result of lqft - as we now show.
Let $P(.)$ denote the Hadamard regularization (or ``partie finie'') of a given integral (see \cite{schwartz} ,pp. 38-43). Then

\underline{Theorem 4.1}

$$
P\bigg(\int_0^{\infty}\!\!E^{vac}(z,d;D)dz\bigg)\;=\;\varepsilon_{f}
\eqno{\rm{(4.5)}}
$$
\vse
\noindent where $\varepsilon_{f}$ is given by (3.17b).

\underline{Proof}.

$$
P\bigg(\!\!\int_{0}^{d}\!\!\!dz\frac{1}{4(2\pi)^2}\frac{1}{z^4}\bigg)\;=\;\frac{1}{4(2\pi)^2}\;\lim_{\epsilon\to0}
\Big[\int_{\epsilon}^{d}\!\!\!dz\;\frac{1}{z^4}\;-\;\frac{1}{3\epsilon^3}\Big]\;=\;
-\frac{1}{12(2\pi)^2}\frac{1}{d^3}
\eqno{\rm{(4.6a)}}
$$
$$
P\bigg(\!\!\int_{0}^{d}\!\!\!dz\frac{1}{4(2\pi)^2}\frac{1}{(d-z)^4}\bigg)\;=\;-\frac{1}{12(2\pi)^2}\frac{1}{d^3}
\eqno{\rm{(4.6b)}}
$$
\vse

\noindent which, inserted into (2.12), yields, after a simple calculation,

$$
P\bigg(\int_0^{\infty}\!\!E^{vac}(z,d;D)dz\bigg)\;=\;
\frac{1}{4(2\pi)^2}\frac{1}{d^3}\times\sum_{m=1}^{\infty} m^{-4}\;=\;
-\frac{1}{2}\frac{\pi^2}{720}\frac{1}{d^3}
\eqno{\rm{(4.7)}}
$$
which equals $\varepsilon_{f}$ by (3.17b).

\underline{Remark 4.1}: (2.12a) shows that the divergence of the energy per unit area remaining after
\underline{renormalization} (1.5) -- i.e., after performing the cancelation between the first integral in (2.3) with the
$m=0$ term in $I_1$ is independent of the ultraviolet cutoff, being only due to the sharp nature of the surface, i.e.,
the use of Dirichlet (or Neumann) b.c. on quantum fields. These are the ``unnatural acts'' referred to by R. Jaffe in
\cite{jaffe}. In the Casimir problem
this divergence is removed by Hadamard regularization, as proved in Theorem 4.1, yielding the correct energy per unit
area. Such was also the finding of Elizalde in \cite{elizalde}: he showed that Hadamard regularization yielded, in models
treated in \cite{jaffe}, the same result for the finite part of the energy density. As seen in (4.5)-(4.7), this procedure identifies precisely the singularities associated to the infinitely thin (sharp) boundary.

\vspace{0.7cm}

\noindent {\bf 5. \ Conclusion}

We have presented in this paper a rigorous method to study the Casimir energy density, and applied it to the massless
scalar field with Dirichlet b.c. on parallel plates. The massive case, as well as other geometries (like cube, sphere)
may in principle be treated by this method, and independence of b.c. has been verified explicitly in each case
(although no general theorem can be invoked).  The finite part $\varepsilon_{f}$ may be derived
in two equivalent ways: 1) with the use of (ultraviolet regularization-dependent) Symanzik surface-renormalization
counterterms, and 2) with no ultraviolet regularization, but applying Hadamard regularization to the
energy density at the singular surfaces . In this connection, we remark that
 (1.6b) becomes a distribution after subtraction of a surface term which may be guessed from (4.1) and (4.2). Let PF denote the Hadamard pseudofunction \cite{schwartz}(pg.41). By (2.12a), (2.13) and the distributional formula \cite{schwartz}(II 1.28)
$$
\frac{d^2}{dz^2} \textrm{PF}(\frac{1}{z^2}) = \textrm{PF}(\frac{1}{z^4})
$$
we obtain the final renormalized one point function
$$
\omega(z) = PF [ \Xi (z- \cdot)]
\eqno{\rm{(5.1)}}
$$
where $\Xi(z,d)$ is defined by
$$
\frac{d^2}{dz^2} \Xi(z,d) = E^{vac}(z,d;D) \mbox{ for }  0<z<d
\eqno{\rm{(5.2)}}
$$
and the dot in (5.1) stands for any point at any of the singular surfaces $z = 0$ and $ z= d$ . Interestingly, this one point function (5.1), which is a constant for periodic b.c., has a singularity structure, with respect to the singular surfaces, of the same nature as the one displayed by the \textbf{two-point} function of a locally covariant quantum field theory \cite{brunetti}, i.e., of the form (see, e.g., \cite{radzikowski}, and \cite{kwald} ):
$$
\omega_{2}(x_1,x_2)=\frac{1}{2\pi^2}\;\textrm{PF}\bigg[\frac{\Delta^{1/2}}{\sigma}\;+\;v\;ln\sigma \bigg]\;+\;w
\eqno{\rm{(5.3)}}
$$
where $\Delta^{1/2}$, $v$ and $w$ are certain smooth functions, and
$\sigma=\sigma (x_1,x_2)$ is the square of the geodesic distance between the points $x_1$ and $x_2$: in (5.2) one has the square of the  distance (here unambiguously defined) between the point $z$ and any one of the singular surfaces. For the two-point function the Hadamard form has a fundamental significance \cite{brunetti}, and it seems worthwhile to inquire on a possibly deeper meaning of this structure in the present context. As remarked by de Witt [37], even the Casimir effect is primarily a problem of Riemannian manifold structure, and, in this sense, locally covariant quantum field theory might place some constraints on the problem.

In conclusion, we have analysed a more realistic model - the massless scalar field enclosed by parallel plates - in the framework of lqft introduced by B.Kay \cite{kay}. Two fundamental methods of asymptotic analysis play a dual role in our treatment. The Poisson summation formula \cite{graham} yields a cutoff independent energy density, which is integrable and homogeneous in the case of periodic b.c., but is not integrable in the region between the plates in the Dirichlet case (also true for Neumann or mixed b.c.). The Euler-MacLaurin summation formula shows the existence, in the asymptotic series for the energy per unit area, of a regularization dependent surface term, which diverges as the cutoff is removed, in the case of Dirichlet b.c.: this term is absent if periodic b.c. are used (remark 3.6). 

We now summarize the final message of the present paper. In addition to local observables such as the energy density (2.12a,b), we propose to regard intensive variables such as the energy per unit area (2.18) as fundamental observables. This is motivated by the fact that in those cases in which the effect is measured, the observed quantity, the pressure, originates from (2.18) by the thermodynamic formula. In applications to Cosmology, for example, one is taken to calculate the Casimir energy of a ``small'' closed region (e.g., a cube) and the pressure by thermodynamics.

Adopting this view, lqft clearly selects periodic b.c.- in this case $\varepsilon_P$ is finite by (2.19b)- rejecting Dirichlet ( Neumann and mixed) b.c., in which case $\varepsilon_D$ diverges by (2.19a). Another way to see this is by the emergence of a RDST in the latter, but not in the former, as shown in section 3. The emergence of the RDST may be seen as a rigorous proof, for the present model, of R.Jaffe's assertion \cite{jaffe} that the renormalized energy is, in general, material dependent, and diverges if a certain parameter controlling the ``sharpness'' of the surface ($\lambda$ in \cite{jaffe}) tends to the value corresponding to an infinitely thin or sharp surface ($\infty$ in \cite{jaffe}). The RDST can certainly be made finite, in the limit where the cutoff is removed, by adopting methods such as those in [32], which allow a macroscopic description of a ``soft'' surface. The latter involves a nonlocal term, as in classical electrodynamics with dielectric materials. A solution of this problem by lqft would probably require a detailed \textbf{microscopic} description of the surface (see, however, [17]).

It may be asked whether the rejection of Dirichlet ( Neumann, mixed) b.c. by lqft is physically plausible, i.e., whether these b.c. are imposed by physics. The only known physical case is the e.m. theory with conductor surfaces (idealized as infinitely thin), where Dirichlet or Neumann b.c. impose themselves by reference to the classical limit, which occurs provided the particles constituting the conductor act collectively in an essentially classical manner [16], but in that case there is a miraculous cancellation ([4],[37]) and no problem arises! The scalar field counterpart to this assertion occurs in the conformally invariant scalar field, as mentioned in the introduction, in which case the nonintegrable terms in (2.12a) disappear, or, in different terms, the RDST in section 3 is absent, as happens in the ordinary scalar quantum field with periodic b.c.. However, as argued in the introduction, the condition of tracelessness, essential for the vanishing of the surface terms, is not met by the (still open) applications of the Casimir effect in Cosmology.

The surface, in Cosmology, is not made out of special materials, but rather used when attempting to restrict the field to some region of space-time which is not reduced to a point, which is obviously required when dealing with quantum fields. The quantity analogous to $\varepsilon$ is the average of the energy density over some closed region. This may lead to a partial justification of the distinguished role of the periodic b.c. : they select just what we conjectured to be the ``field component'' of the vacuum energy ( equal to $\varepsilon_{f}$ in (3.17b) up to a factor) which, moreover, may be positive in the massive case [36], and yields, through the renormalization prescription (1.5), a \textbf{homogeneous} energy density (2.19b) of the type which is required in general relativity. Note that, in the case of Dirichlet b.c., even the finite part of the energy density (5.1) is \textbf{not} homogeneous. A basic open problem is the dependence on the region's dimension ($d^{-3}$ in (2.19b), $L^{-4}$ for a cube of side $L$ [6]): taking $L$ of order of the ``radius of the Universe'' leads to an absurdly small value of the dark energy density [12]. 

Acknowledgements:
We are greatly indebted to the three referees, most particularly referee1, whose comments considerably clarified the conceptual basis of this paper.

\vspace{0.7cm}

\pagebreak

\end{document}